\documentclass[preprint,11pt]{aastex}

\newcommand{\lam}{$\lambda$}

\newcommand{\ecss}{erg~cm$^{-2}$~s$^{-1}$~sr$^{-1}$} % erg/cm2/s/sr
\newcommand{\kms}{km~s$^{-1}$}
\newcommand{\as}{${^\prime}{^\prime}$}

\renewcommand{\ion}[2]{#1\,{\sc #2}}

\newcommand{\hinode}{\emph{Hinode}}

\def\ionx[#1 #2]{#1\,{\sc #2}}

\title{Dark jets in solar coronal holes} 

\author{Peter R. Young}

\affil{College of Science, George Mason University, 4400 University
  Drive, Fairfax, VA 22030} 

\begin{document}

\begin{abstract}
A new solar feature termed a \emph{dark jet} is identified from
observations of an extended solar coronal hole that was continuously monitored for over 44
hours by the EUV Imaging Spectrometer on board the \hinode\
spacecraft in 2011 February 8--10. Line-of-sight velocity  maps derived from  the  coronal
\ion{Fe}{xii} \lam195.12 emission line, formed at 1.5~MK,  revealed a number of
large-scale, jet-like structures that showed significant blueshifts. The
structures had either weak or no intensity signal 
in 193~\AA\ filter images from the Atmospheric Imaging Assembly on board
the Solar Dynamics Observatory, suggesting that the jets are
essentially invisible to imaging instruments. The
dark jets are rooted in bright points and occur both within the
coronal hole and at the quiet Sun--coronal hole boundary. They exhibit
a wide range of shapes, from narrow columns to fan-shaped structures,
and sometimes  multiple jets are seen close together.
A detailed study of one dark jet showed line-of-sight speeds increasing
along the jet axis from 52 to 107~\kms\ and a temperature of
1.2--1.3~MK. The low intensity of the jet was due either to a small
filling factor of 2\%\ or to a curtain-like morphology. From the
HOP~177 sample, dark jets are as common as regular coronal hole jets,
but their low intensity suggests a mass flux around two orders of
magnitude lower.
\end{abstract}


\section{Introduction}

Coronal jets are a striking feature of solar coronal hole observations
obtained at X-ray or extreme ultraviolet (EUV) wavelengths
\citep{shimojo96,cirtain07,nistico09}. They are identified through a
transient, collimated structure that appears in image sequences and
thus, by definition, have an enhanced intensity over their
background. In this work we show examples of jets that are essentially
invisible in EUV image sequences, but have a clear signature in
Dopplergrams derived from an EUV emission line. We refer to these
events as \emph{dark jets}.

Jets are  a fundamental type of energy-release process on the Sun,
with a relatively simple observational signature. As such they have
been the subject of extensive theoretical study ranging from the early
2D simulations of \citet{shibata86} and \citet{yokoyama95}, to more
recent 3D simulations of \citet{miyagoshi03}, \citet{pariat09} and 
\citet{moreno13}. The jets represent an outflow of plasma that is
believed to be driven by magnetic field evolution, and  flux emergence
is the most commonly-modeled scenario \citep{yokoyama95,moreno13}
although observations suggest that flux cancellation often leads to
jets \citep{liu11,young14-pasj,young14-sp}.

The modern observatories \hinode\ and the Solar Dynamics Observatory
(SDO) have led to many new jet studies, particularly for coronal hole
jets (CHJs). The X-Ray Telescope (XRT) on board \hinode\ sees many more
CHJs than previous X-ray instruments \citep{cirtain07,savcheva07},
principally due to enhanced sensitivity at lower
temperatures. Coverage has also been expanded by the Atmospheric
Imaging Assembly (AIA) on board SDO which obtains full-disk solar
images in a number of EUV filters at a continuous, high time-cadence.
Examples of CHJs from AIA were presented by \citet{shen11} and \citet{hong13},
and the present work follows on from the work of \citet{young14-pasj,young14-sp}
who combined SDO imaging data with
spectroscopic data from the EUV Imaging Spectrometer (EIS) on board
\hinode.

The present paper is structured as follows. Sect.~\ref{sect.obs}
describes the observations; Sect.~\ref{sect.eg} presents three
examples of dark jets, demonstrating their lack of an intensity
signature; Sect.~\ref{sect.jet4} provides a detailed analysis of one
of the dark jets; and results are summarized in Sect.~\ref{sect.summary}.

\section{Observations}\label{sect.obs}

The dark jets discussed in the present work were identified from a
data-set obtained through 
Hinode Operation Plan (HOP) No.~177, which was run over a 44.4 hour period
from 2011 Feburary 8 10:22~UT to February 10 06:47~UT. 
The jets are identified from Hinode/EIS data and SDO/AIA images are
used for comparison.
AIA is described by \citet{aia} and 
we mostly use images from the 193~\AA\ filter, which we refer to as
``A193''. This filter is dominated by emission from \ion{Fe}{xii} in
most conditions \citep{odwyer10} and so is the best comparison with
the EIS \ion{Fe}{xii} \lam195.12 emission line. The images are
obtained at a 12~second cadence and we bin
groups of five images together in order to boost signal-to-noise,
giving a cadence of 1~minute. For studying the evolution of the bright
points at the bases of the dark jets, we consider A193 images for the period $\pm
20$~minutes either side of the time when EIS observed the jets. 
The EIS
instrument \citep{culhane07} performed 43 raster scans with the study Large\_CH\_Map,
which performed a  scan over a field-of-view 179\arcsec\
$\times$ 512\arcsec\ with the 2\as\ slit at 3\as\ step sizes and with 
60~second exposure times.

The EIS data were calibrated using the standard options recommended in
the EIS data-analysis
guide\footnote{\url{http://solarb.mssl.ucl.ac.uk:8080/eiswiki/Wiki.jsp?page=EISAnalysisGuide}.} 
and the \ion{Fe}{xii} \lam195.12 emission line was fit with a Gaussian
function at each spatial pixel in the rasters. This line was selected
as it is the strongest coronal emission line, in terms of detected
photons,  in EIS coronal hole spectra
 due to the high instrument sensitivity at this wavelength.
From the line fits,
images of intensity, line-of-sight (LOS) velocity and line width were
created. The velocity maps revealed 35 spatial features within the
coronal hole that exhibited
blue-shifts of at least 15~\kms\ over an extended spatial area. Each feature was present in only a
single EIS raster, and so their lifetimes were less than 62~minutes
(the cadence of the rasters). Twenty-four of the features could be
classified as jets in that they exhibited collimated structures
aligned roughly radially from Sun center. 
AIA 193~\AA\ images were studied in
order to identify dynamic phenomena related to the jets. For 13 events
we could identify collimated intensity structures that matched the
morphology of the EIS velocity structures, and two examples were studied
by \citet{young14-pasj,young14-sp}. There remained 11 events
that showed collimated, jet-like structure in the velocity maps but
for which there was not a clear intensity signal in the A193
images. Since jets are typically identified from image sequences
through their intensity enhancement then we refer to jets identified
through a velocity signature as ``dark jets''. 

The locations of the 35 blue-shifted features found from the EIS
dopplergrams are plotted on an A193 image in Figure~\ref{fig.all}. The
image was obtained by averaging 10 consecutive A193 images between
February~8 23:59~UT and February~9 00:01~UT, and the locations of the
EIS events have been corrected for the solar rotation. The blue line on
Figure~\ref{fig.all} shows the coronal hole boundary as determined by
the SPOCA code \citep{delouille12}, made available through the
Heliophysics Event Registry \citep{hurlburt12}. We note that the
coronal hole has a significant amount of mixed polarity, leading to
quite large bright points such as the bright feature at position
$(-380,-670)$. We believe that this is a bright point within the
coronal hole rather than a quiet Sun region that intrudes into the
coronal hole. This is based on the low A193 intensity seen all around
the bright point. As will be discussed in the following section, this
bright point produced a number of dark jets.

Further details on the full range of jet events identified from
HOP~177 are available at the website
\url{http://pyoung.org/jets/hop177}, which also contains movies for
all of the dark jets discussed here.
We proceed in the next section to present
three examples of dark jets.

\section{Dark jet examples}\label{sect.eg}

\begin{figure}[h]
\epsscale{0.6}
\plotone{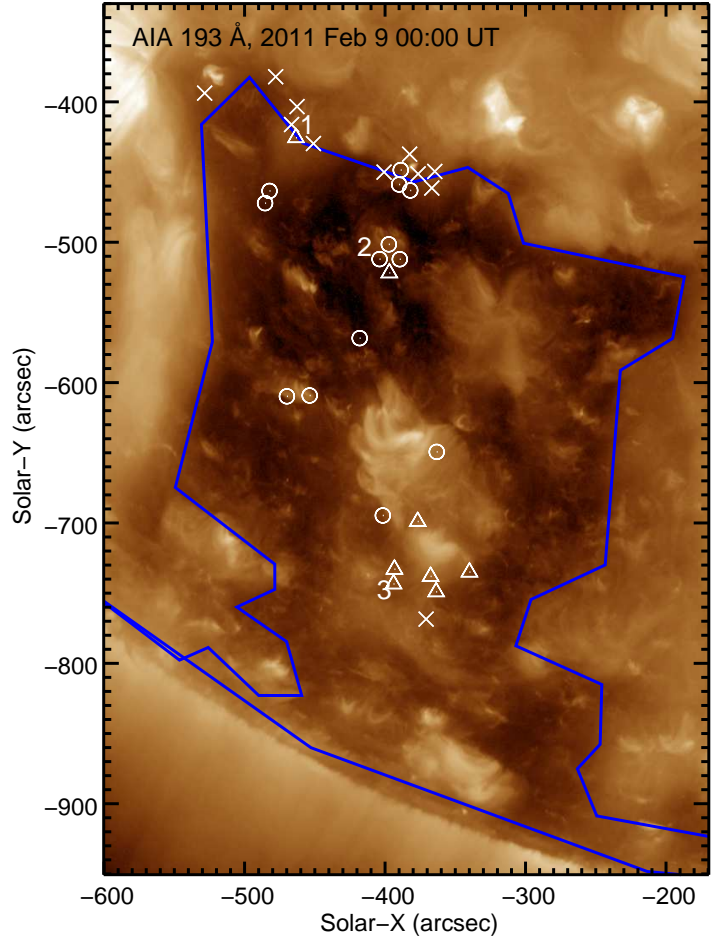}
\caption{An A193 image from 00:00~UT on February~9, with a logarithmic
intensity scaling applied. The symbols and numbers mark the location
of 35 blue-shifted features identified from the HOP~177
data-set. Their locations have been differentially-rotated to match
the image time. Coronal hole jets are marked with circles; dark jets
are marked with triangles or numbers; other blue-shifted features are
marked with crosses. The blue line indicates the coronal hole boundary
determined by the SPOCA code \citep{delouille12}.}
\label{fig.all}
\end{figure}

\begin{figure}[h]
\epsscale{1.0}
\plotone{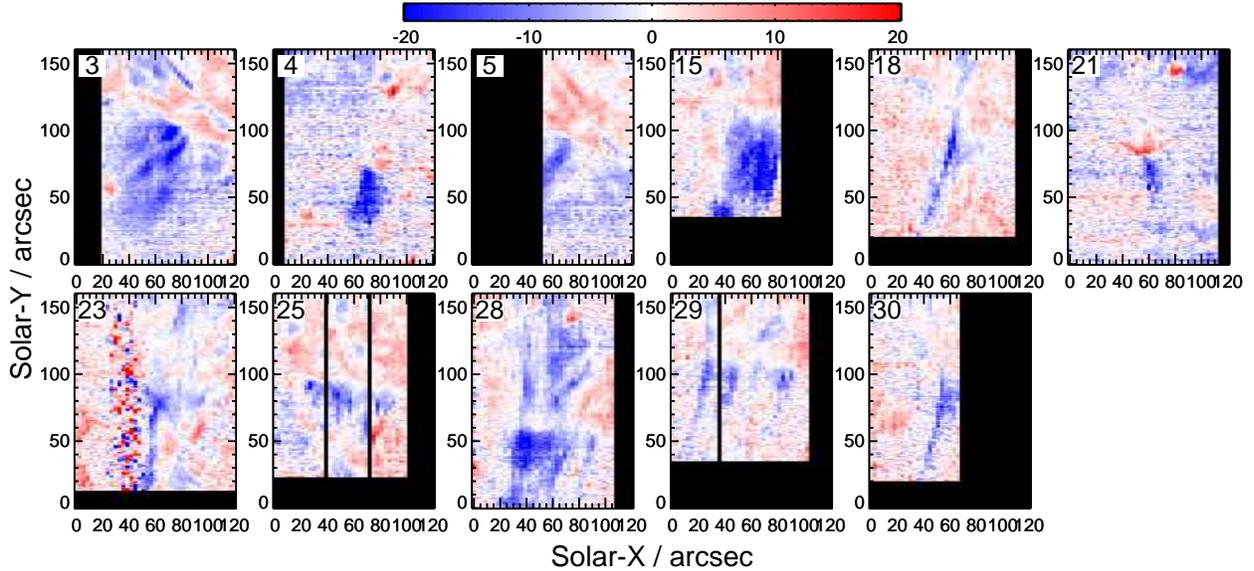}
\caption{\ion{Fe}{xii} \lam195.12 dopplergram images of all of the dark jets identified from
  the HOP~177 data-set, with indices from Table~\ref{tbl.list}. LOS
  velocities in the range $-20$ to 
  $+20$~\kms\ are shown, and each image has a size of $120\times
  150$~arcsec$^2$. }
\label{fig.ims}
\end{figure}

Figure~\ref{fig.all} shows the locations of all of the 35 blue-shifted
features found from the HOP~177 data-set. Circles show the locations
of jets that have an A193 signature, and crosses show features that were
not classified as jets due to a lack of a collimated structure. Dark
jets are identified with triangles, except for the three jets labeled
1, 2 and 3. These three jets are discussed in more detail below. 
Of the 11 dark jets, two came from a location on the coronal hole
boundary, two came from a small bright point, and seven came from a
large feature that we call a bright point complex.
Details of the
times and positions of these jets are given in
Table~\ref{tbl.list}. Each of the 35 blue-shifted features were assigned an
index number ordered according to the time the features were observed, and the indices
of the 11 dark jets are given in Table~\ref{tbl.list}.
Dopplergrams derived from 
\ion{Fe}{xii} \lam195.12 for each of the dark jet locations are shown in Figure~\ref{fig.ims}.
These
images -- derived by fitting a Gaussian to \lam195.12 and converting
the centroid to a velocity by comparing with the average centroid over
the whole raster -- show a wide range of morphology. Events 18,
23 and 30
show long, narrow structures; events 3 and 15 show broad jets;
events 25 and 29
actually consist of multiple jets close together; and event 28 has a very
complex structure that extends over 100\as\ in length. Below we
consider three of the events, and compare the EIS features with A193
image sequences.

\begin{deluxetable}{cccccc}
\tablecaption{EIS dark jets from HOP~177.\label{tbl.list}}
\tablehead{
  \colhead{Index} &
  \colhead{Date} &
  \colhead{Time} &
  \colhead{RN\tablenotemark{a}} &
  \colhead{(X,Y)} 
  % \colhead{XRT} &
  % \colhead{SOT} &
}
\startdata
  3 &    8-Feb &   13:13 &   3 & ($-530$,$-420$)\\
  4 &          &   15:58 &   6 & ($-465$,$-505$)\\
  5 &          &   16:16 &   6 & ($-515$,$-425$)\\
 15 &          &   23:59 &  14 & ($-340$,$-735$)\\
 18 &    9-Feb &   01:07 &  15 & ($-390$,$-745$)\\
 21 &          &   04:12 &  18 & ($-370$,$-515$)\\
 23 &          &   07:23 &  21 & ($-365$,$-730$)\\
 25 &          &   08:18 &  22 & ($-335$,$-735$)\\
 28 &          &   10:25 &  24 & ($-330$,$-695$)\\
 29 &          &   11:26 &  25 & ($-320$,$-745$)\\
 30 &          &   13:42 &  27 & ($-350$,$-740$)\\

\enddata
\tablenotetext{a}{EIS raster number (between 1 and 43).}
\end{deluxetable}

\begin{figure}[t]
\epsscale{1.0}
\plotone{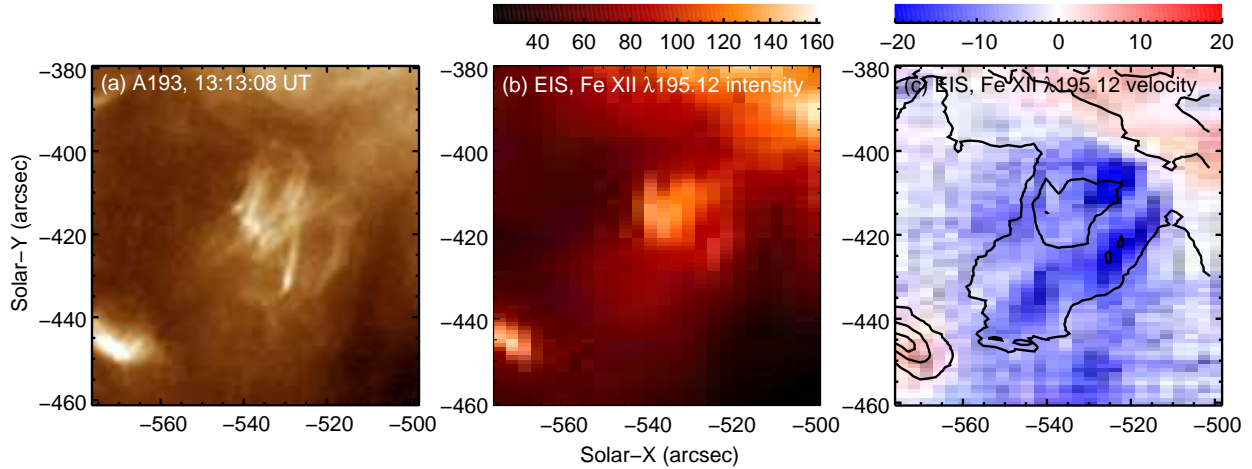}
\caption{Images of a dark jet observed at 13:13~UT on February
  8. Panel (a) shows an A193 image averaged from five consecutive
  images obtained over 60~seconds. Panels (b) and (c) show intensity
  (units: \ecss) and LOS velocity (units: \kms) images derived from
  performing Gaussian fits to \ion{Fe}{xii} \lam195.12. }
\label{fig.eg1}
\end{figure}

\begin{figure}[t]
\epsscale{0.7}
\plotone{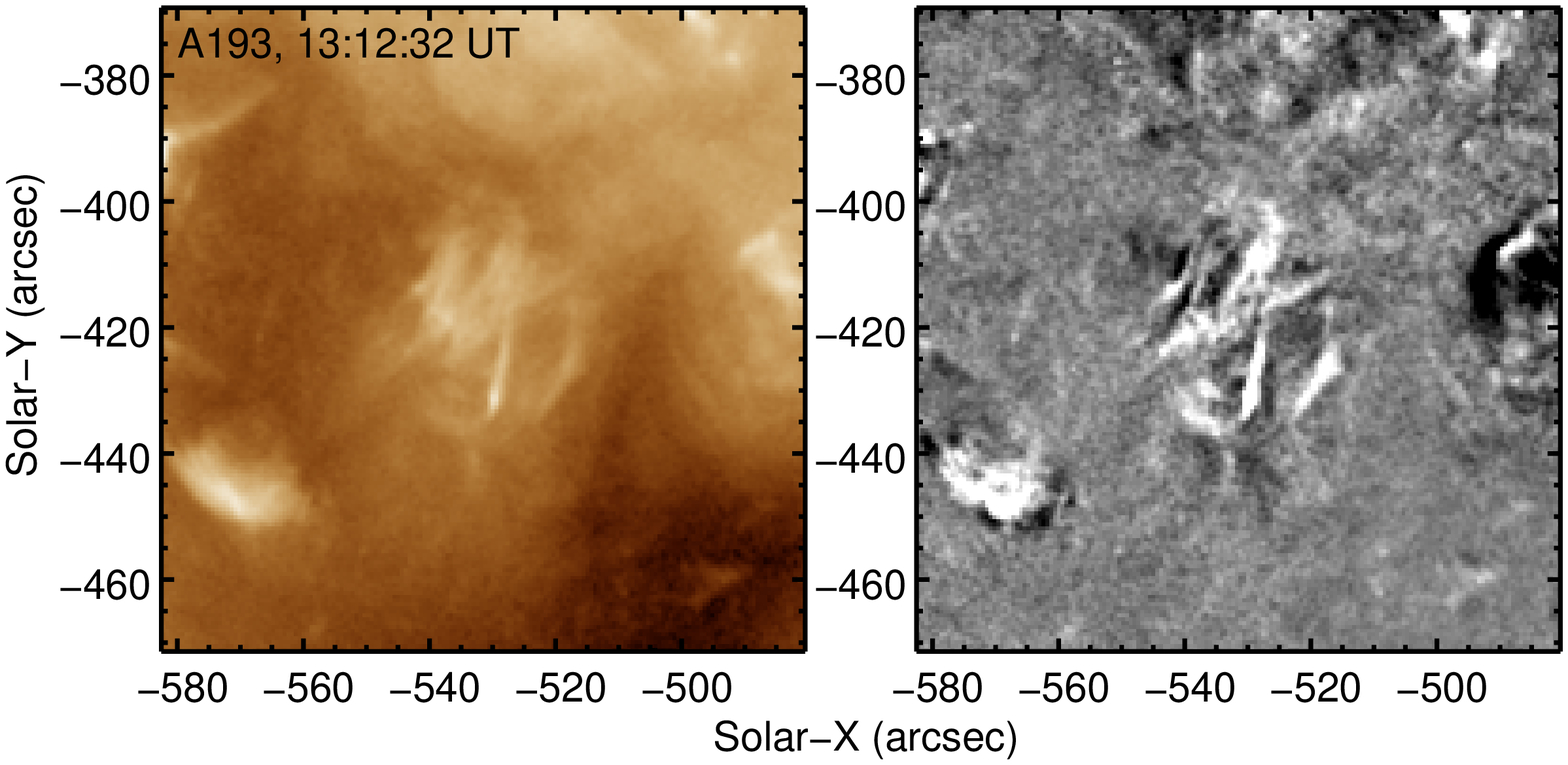}
\caption{A single image from a movie available in the online edition
  of the journal, showing the evolution of the bright point that gives
  rise to the dark jet shown in Figure~\ref{fig.eg1}. The left panel
  shows a A193 image with a logarithmic 
intensity scaling applied, and the right panel shows a difference
movie where the mean image over the 40-image sequence is subtracted
from the left-hand image.}
\label{fig.mov1}
\end{figure}

Figure~\ref{fig.eg1} shows images from a dark jet that EIS rastered
over at around 13:13~UT on February 8. It occurred near the coronal
hole boundary (which can be seen in the top-right corner of the
images). A bright point can be seen in the \lam195.12 intensity image
(Figure~\ref{fig.eg1}b), and the A193 image (Figure~\ref{fig.eg1}a)
resolves the bright point into a number of small loops. The EIS
velocity image (Figure~\ref{fig.eg1}c) shows extensive blue-shifts at
the bright point and extending radially away from it. The A193
1-minute cadence movie and the difference image movie
(Figure~\ref{fig.mov1}) show that the
bright point is quite dynamic, but there is no clear evidence of a jet
or jets
coming from the bright point that could be responsible for the
blue-shifts seen in the EIS data.
The bright point was present for about two days from 12~UT on February
7, and the A193 emission remained rather weak and diffuse over this
period. 

\begin{figure}[t]
\epsscale{1.0}
\plotone{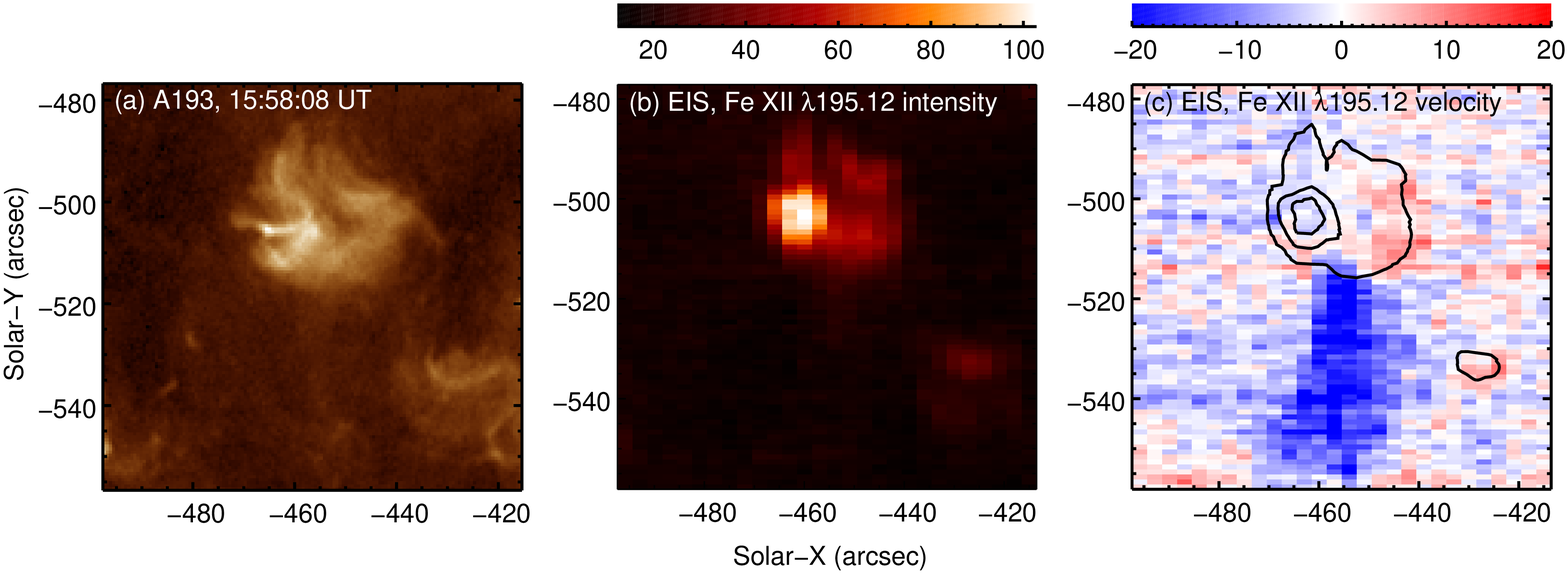}
\caption{Images of a dark jet observed at 15:58~UT on February
  8. Panel (a) shows an A193 image averaged from five consecutive
  images obtained over 60~seconds. Panels (b) and (c) show intensity
  (units: \ecss) and LOS velocity (units: \kms) images derived from
  performing Gaussian fits to \ion{Fe}{xii} \lam195.12.}
\label{fig.eg2}
\end{figure}

\begin{figure}[t]
\epsscale{0.7}
\plotone{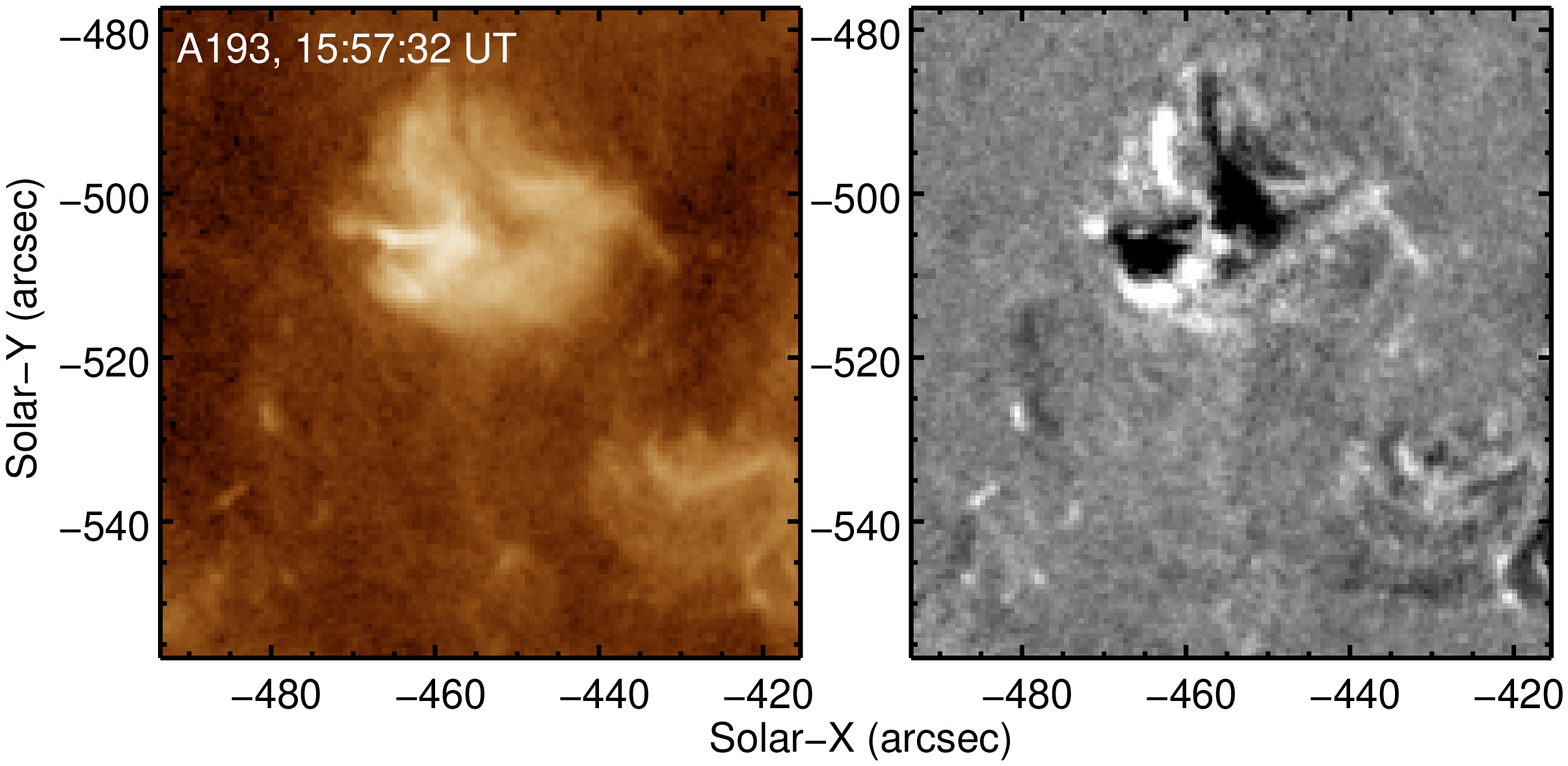}
\caption{A single image from a movie available in the online edition
  of the journal, showing the evolution of the bright point that gives
  rise to the dark jet shown in Figure~\ref{fig.eg2}. The left panel
  shows a A193 image with a logarithmic 
intensity scaling applied, and the right panel shows a difference
movie where the mean image over the 40-image sequence is subtracted
from the left-hand image.}
\label{fig.mov2}
\end{figure}

The second dark jet example comes from a small bright point that is
isolated within the coronal hole. The bright point was present for
around two days, and it gave rise to a number of jets including the
blowout jet described in \citet{young14-sp}, which was caused by the
cancelation of the main polarities of the bright point leading to its
eventual disappearance. The blowout jet occurred at 09:00~UT on
February 9, and the dark jet presented here was observed  at 15:58~UT
on February 8.  Figure~\ref{fig.eg2} shows a clear jet-like feature in
the velocity map, but with no counterpart in the A193 image or EIS
\lam195.12 intensity image. Neither the A193 image movie nor the
difference movie (Figure~\ref{fig.mov2}) show any structures extending away from the bright
point that can be identified with the EIS velocity feature. We note
that another dark jet from the same bright point was captured by EIS
at 04:12~UT on February 9.

\begin{figure}[t]
\epsscale{1.0}
\plotone{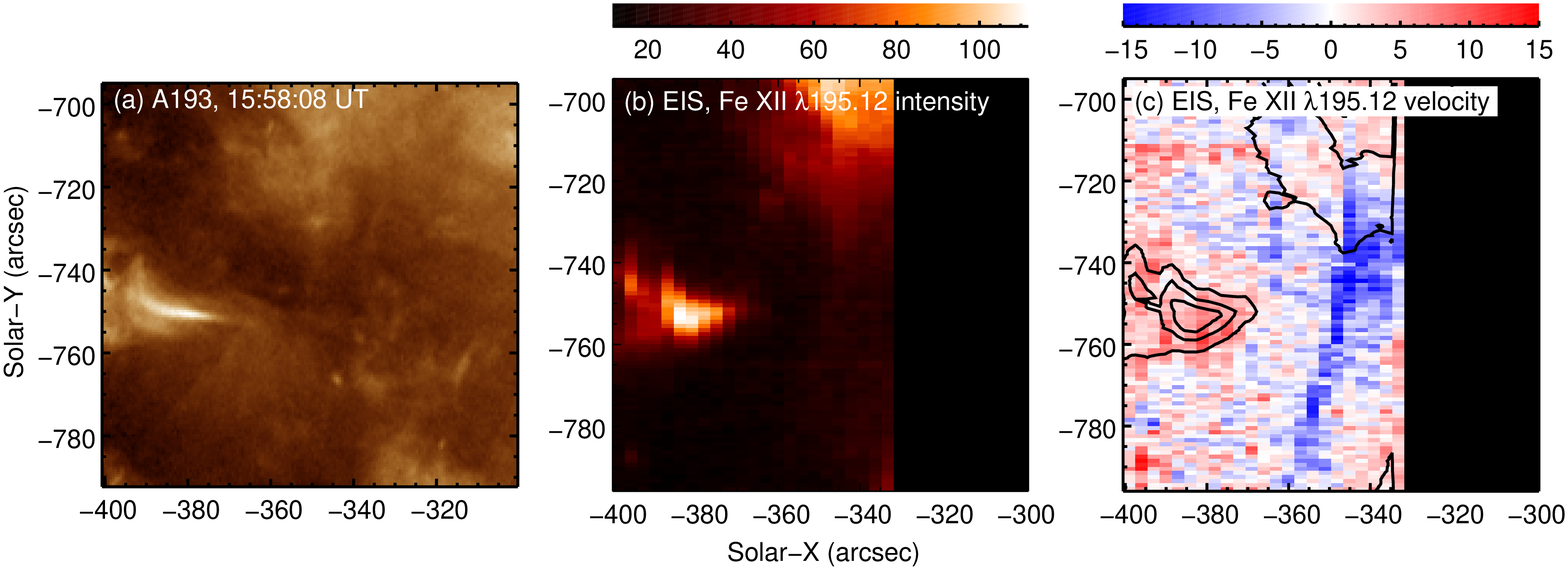}
\caption{Images of a dark jet observed at 13:42~UT on February
  9. Panel (a) shows an A193 image averaged from five consecutive
  images obtained over 60~seconds. Panels (b) and (c) show intensity
  (units: \ecss) and LOS velocity (units: \kms) images derived from
  performing Gaussian fits to \ion{Fe}{xii} \lam195.12.}
\label{fig.eg3}
\end{figure}

\begin{figure}[h]
\epsscale{0.3}
\plotone{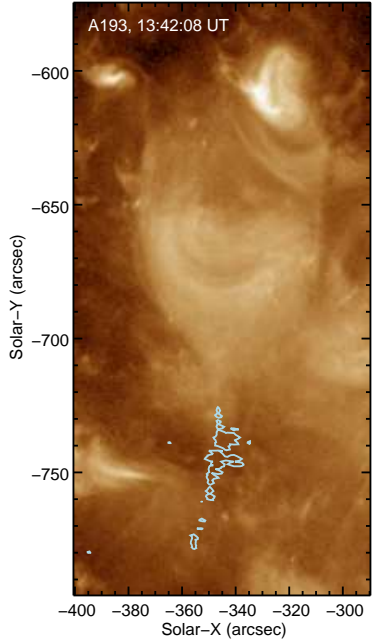}
\caption{An A193 image from February 9 13:42:08~UT with a logarithmic
  intensity scaling. The light blue contours show EIS Doppler
  velocities of $-10$~\kms, showing the location of the dark jet from Figure~\ref{fig.eg3}. }
\label{fig.eg3-big}
\end{figure}

\begin{figure}[t]
\epsscale{0.7}
\plotone{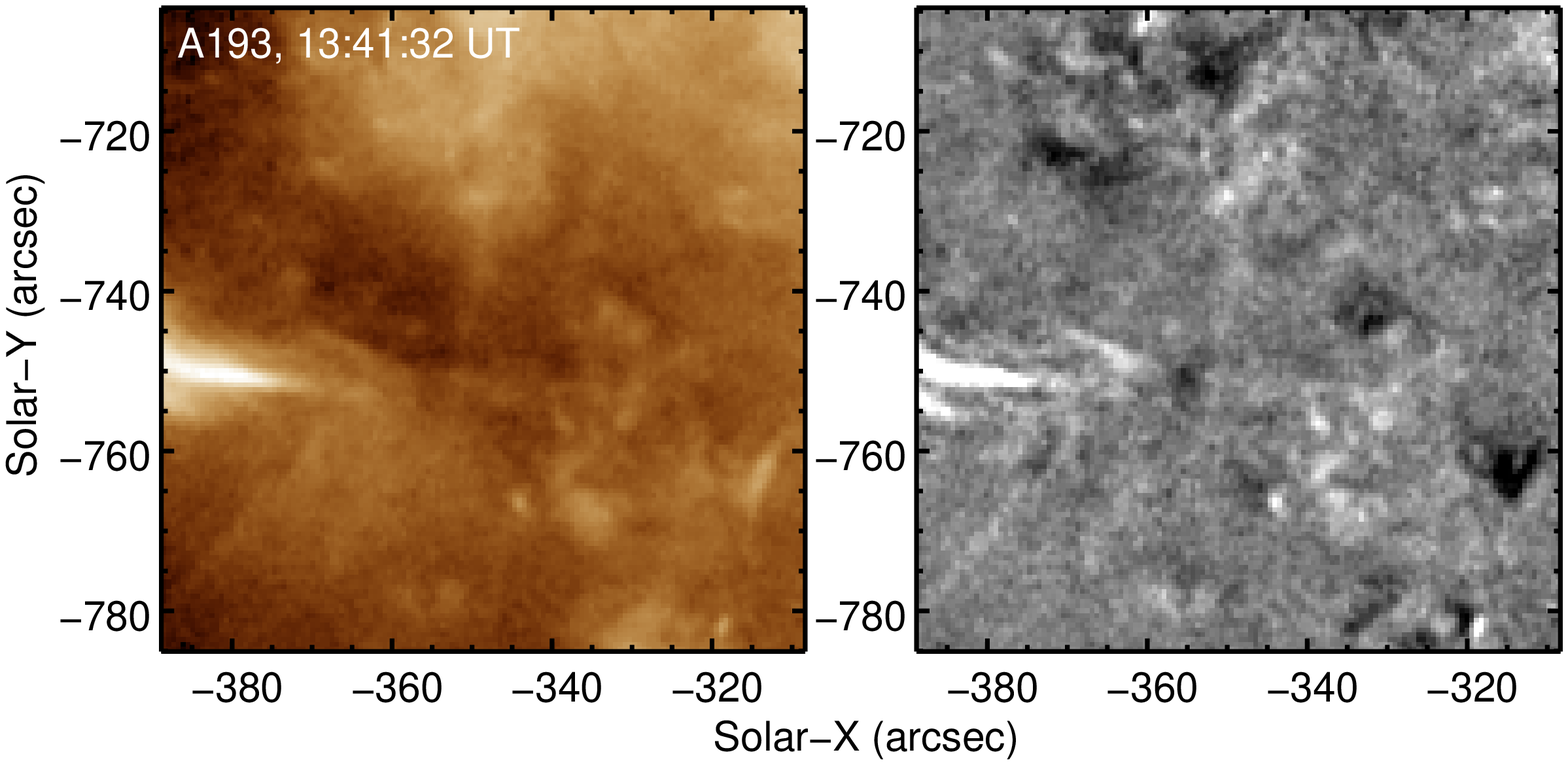}
\caption{A single image from a movie available in the online edition
  of the journal, showing the evolution of the bright point that gives
  rise to the dark jet shown in Figure~\ref{fig.eg3}. The left panel
  shows a A193 image with a logarithmic 
intensity scaling applied, and the right panel shows a difference
movie where the mean image over the 40-image sequence is subtracted
from the left-hand image.}
\label{fig.mov3}
\end{figure}

Figure~\ref{fig.eg3} shows the third example of a dark jet, which is
evident as a blue collimated structure in panel c. The EIS raster is
truncated on the right-hand side due to lost telemetry packets during
the observation. The jet originates in a region of complex morphology,
and in Figure~\ref{fig.eg3-big} we show a larger A193 field-of-view
with the EIS jet location identified. There is an intense bright point
at $(-325,-610)$ with larger, more diffuse loops related to this bright
point in the region $Y=-700$ to $-620$. There is another patch of emission
around $(-355,-720)$ that may be a distinct bright point. It is not
clear if the jet is from this bright point or if it has some
connection to the large, diffuse loops. The A193 movie and
difference-movie (Movie~\ref{fig.mov3}) do not show evidence for an
intensity structure that matches the EIS velocity jet.

\section{The February 8 15:58~UT jet}\label{sect.jet4}

In this section we take a closer look at the dark jet that occurred on
February 8 at 15:58~UT (Figures~\ref{fig.eg2}, \ref{fig.mov2}). The jet is 
chosen as it has the simplest geometry, belonging to an isolated bright
point in the coronal hole.

\begin{figure}[h]
\epsscale{1.0}
\plotone{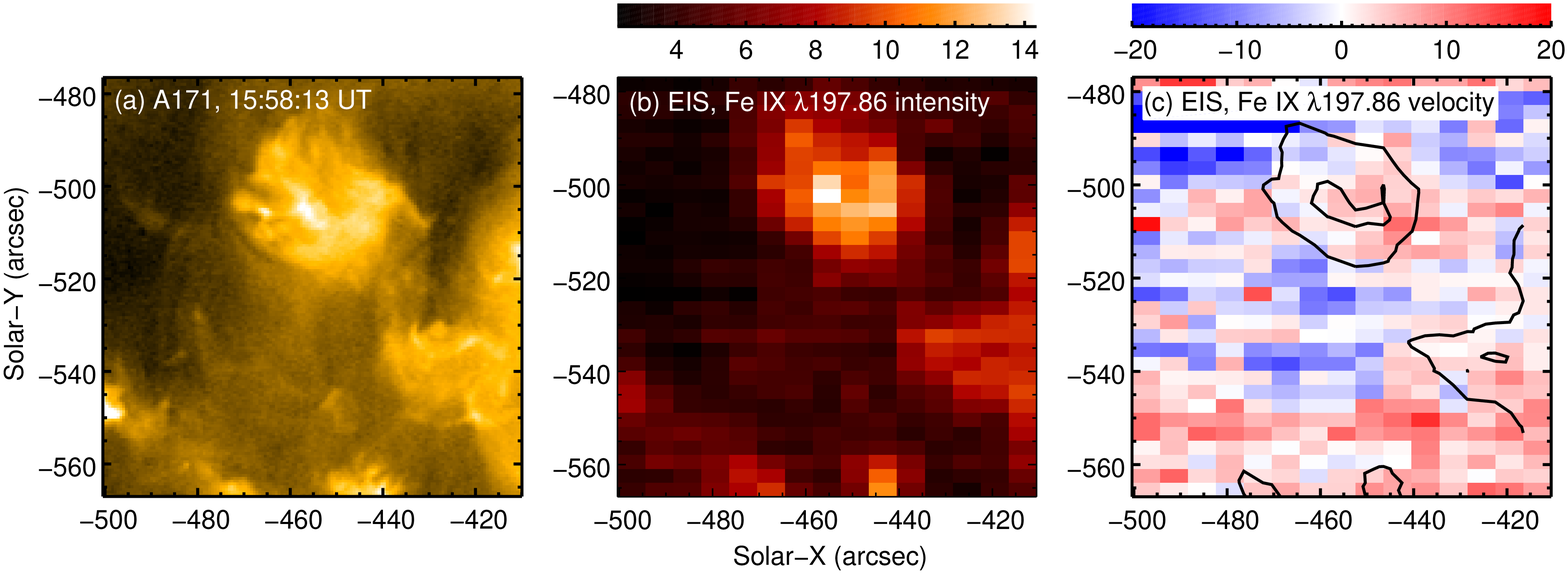}
\caption{ Panel (a) shows an A171 image averaged from five consecutive
  images obtained over 60~seconds for the dark jet on February 9,
  15:58~UT (Figure~\ref{fig.eg2}). Panels (b) and (c) show intensity
  (units: \ecss) and LOS velocity (units: \kms) images derived from
  performing Gaussian fits to \ion{Fe}{ix} \lam197.86}
\label{fig.fe9}
\end{figure}

Firstly we note that the morphology of the blue feature in the velocity map
(Figure~\ref{fig.eg2}c) is suggestive of a coronal hole plume, and an
A171 image (Figure~\ref{fig.fe9}a) does show some diffuse emission extending
from the bright point that could be plume plasma. (Polar plumes are
known to have temperatures close to the formation temperature of the
\ion{Fe}{ix} \lam171.1 emission line.) Therefore the \lam195.12
outflow could represent outflowing plume plasma. This can be
discounted, however, because the \lam195.12 velocity feature is only seen in a
single raster, implying a lifetime of $< 62$~minutes, whereas plumes
are known to have lifetimes of the order of a day \citep[see review
of][]{wilhelm11}. In addition, since plumes have temperatures of
0.7--1.1~MK \citep{wilhelm11} then the implied outflowing
plasma would be best seen in emission lines formed at this
temperature. EIS observes \ion{Fe}{ix} \lam197.86 and
Figures~\ref{fig.fe9}b,c show intensity and velocity images formed
from this line. Since it is much weaker than \lam195.12 it was
necessary to rebin the data into $2\times 3$ pixels, but
Figure~\ref{fig.fe9}c clearly shows that there is no velocity feature
to compare with that of \lam195.12. Note that the statistical uncertainties on
the \lam197.86 velocities are $\approx \pm 5$~\kms.

Figure~\ref{fig.mov2} demonstrated that an intensity signature of the
dark jet could not be seen in an A193 image sequence. The remaining AIA EUV
filter image sequences (94, 131, 171, 211,
304 and 335~\AA)
were also checked, but no signature was found. Although \hinode\ X-Ray
Telescope data were
available for this event, the filters
(thin-beryllium and titanium-poly) were not suitable for studying the
faint jet emission, and only the bright point could be seen. Previous
XRT jet studies \citep{cirtain07,savcheva07} used the aluminium-poly
filter which has a better response to low temperature plasmas.

\begin{figure}[h]
\epsscale{0.7}
\plotone{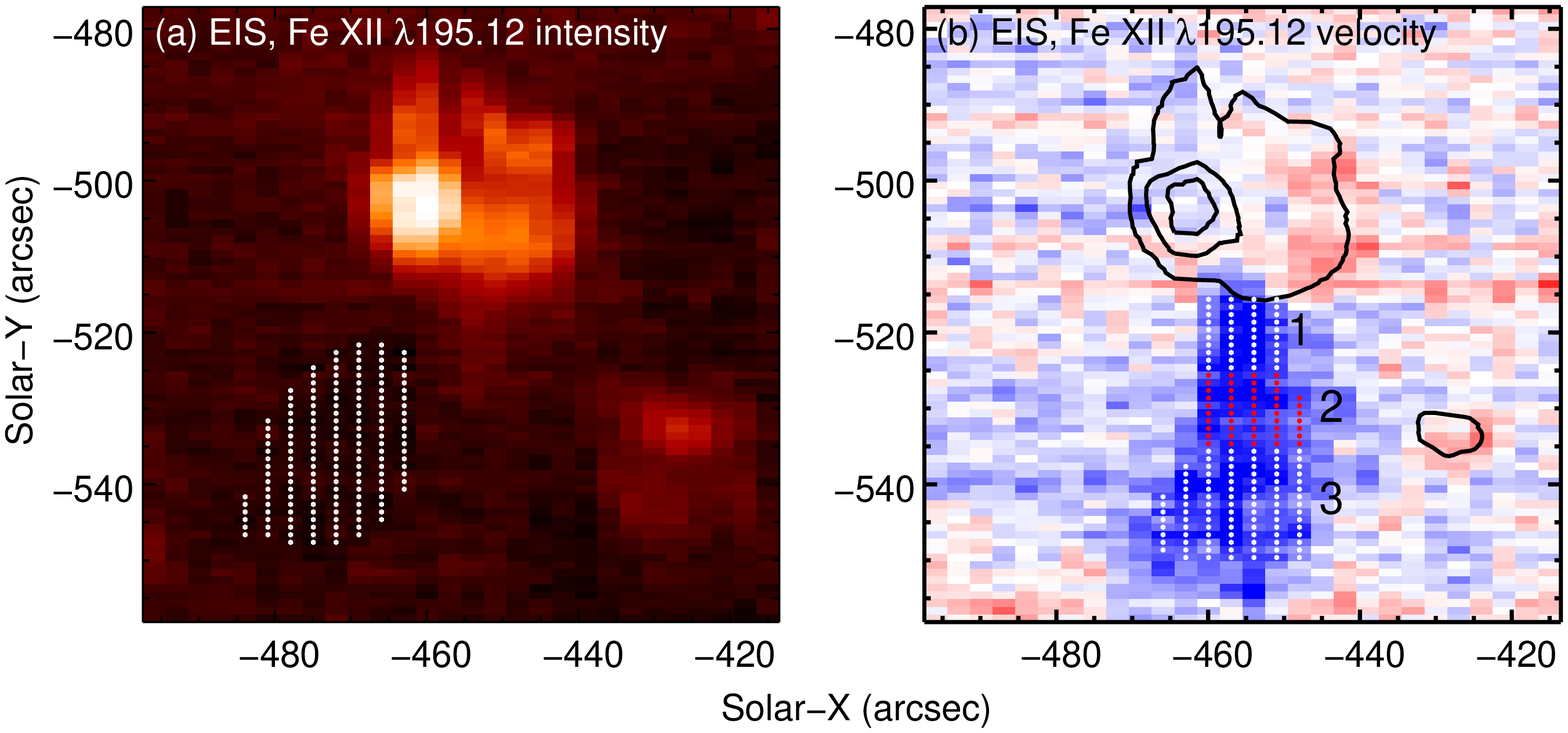}
\caption{The left panel shows an EIS intensity image from
  \ion{Fe}{xii} \lam195.12 for the February 8 15:58~UT jet. The region
selected as the coronal hole background is indicated. The right panel
shows the \ion{Fe}{xii} \lam195.12 Dopplergram for the same jet, and
three regions along the jet's axis are indicated.}
\label{fig.sub}
\end{figure}

\begin{figure}[h]
\epsscale{0.7}
\plotone{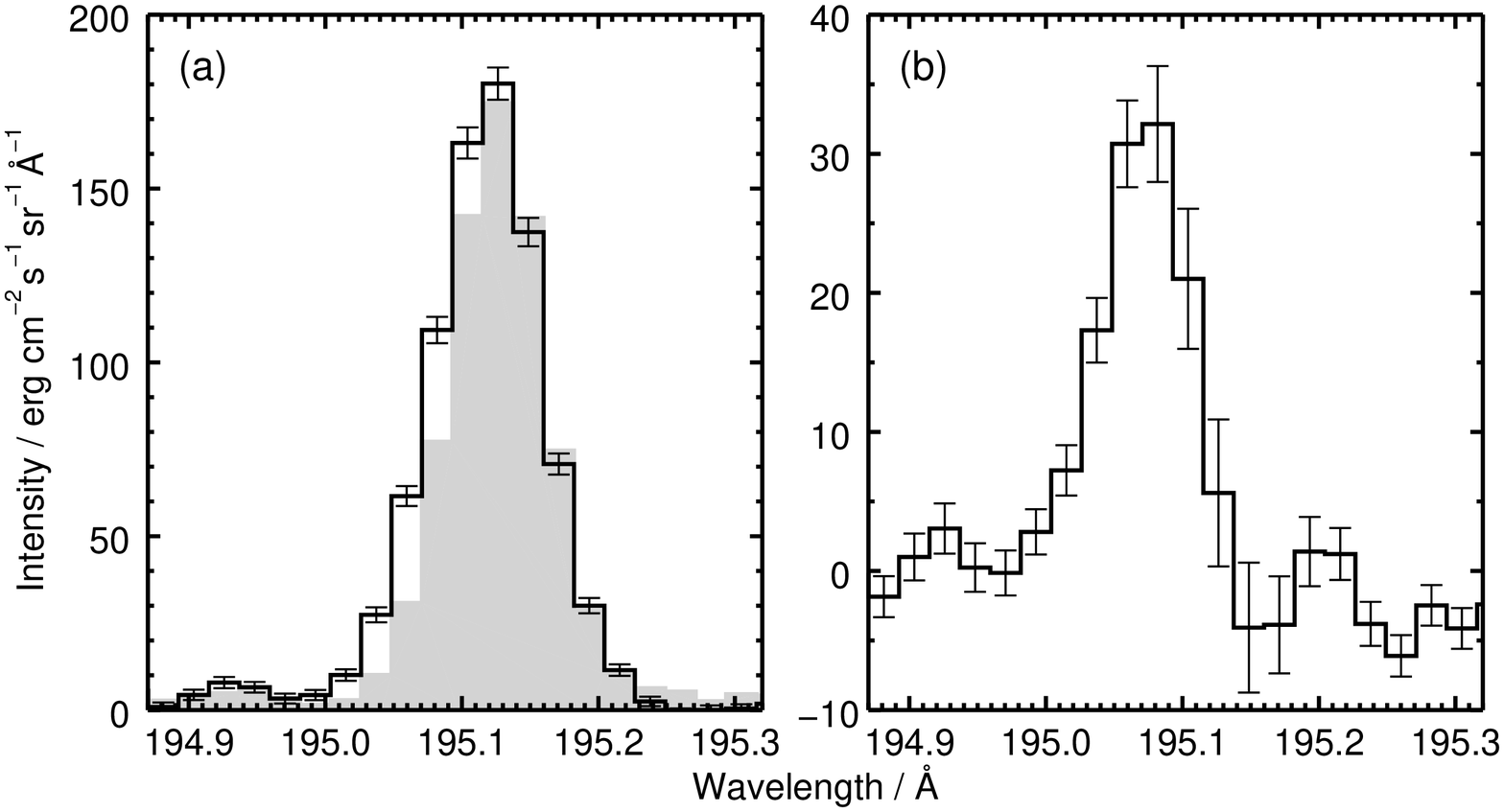}
\caption{The left panel shows the averaged EIS spectrum from jet
  region 2 (Figure~\ref{fig.sub}b), with the spectrum from the coronal
hole background region (Figure~\ref{fig.sub}a) over-plotted in
gray. The right panel shows the spectrum derived by subtracting the
background spectrum from the jet spectrum. Error bars are derived from
Poisson statistics.}
\label{fig.subspec}
\end{figure}

Inspection of \lam195.12 profiles shows that the dark jet blueshift
seen in Figure~\ref{fig.eg2}c is due to an asymmetry in the profile caused by
extra emission on the short-wavelength side of the line. As the jet
occurs within the coronal hole background, then the line profile
is a composite of the background and jet emission. To isolate the jet
component we subtract out the background component with a technique
illustrated in Figures~\ref{fig.sub} and \ref{fig.subspec}. Four
spatial regions were identified: one in the coronal hole background
neighboring the jet (Figure~\ref{fig.sub}a), and three along the axis
of the jet (Figure~\ref{fig.sub}b). 
Averaged spectra from each region were
obtained using the IDL routine EIS\_MASK\_SPECTRUM. The keyword /SHIFT
was applied, which shifts the spectrum at each spatial pixel onto a
common wavelength scale. In particular this accounts for offsets that
arise due to the thermal drift and slit tilt found in the EIS data
\citep{kamio10}. Figure~\ref{fig.subspec}a shows the spectrum from jet
region 2, with the background region spectrum overplotted. The
background-subtracted spectrum for jet region 2 is shown in
Figure~\ref{fig.subspec}b, clearly revealing a Gaussian-shaped
feature that represents the jet plasma. The intensity, width and velocity
of this feature for the jet regions 1--3 are given in
Table~\ref{tbl.fits}. 
Note that the velocity of this component was derived by assuming that
\lam195.12 in the background spectrum is at rest.

The jet velocities are
derived relative to the position of \lam195.12 in the background
spectrum. The velocity of the jet component increases along the length
of the jet from $-52$ to $-104$~\kms, a behavior consistent with the
findings of \citet{young14-pasj,young14-sp} 
for two blowout jets and which suggests that
material continues to be accelerated along the jet. 
The intensities assume the original
laboratory radiometric calibration of EIS \citep{lang06}. The revised
calibration of \citet{dz-cal} does not modify the \lam195.12
intensities, whereas the revised calibration of \citet{warren14}
reduces the intensities by a factor of 0.85. The line widths have
been corrected for the EIS instrumental width \citep{eis_sw7}, and are
larger than the thermal width at $\log\,T=6.2$ (the temperature of
maximum emission of \lam195.12), which is 23~m\AA,
although there is no clear pattern when comparing with the background
width.

Density measurements of the jet plasma are not possible as
the density-sensitive line \lam186.88 (actually a blend of two
\ion{Fe}{xii} lines at 186.85 and 186.89~\AA) could not be measured in
the subtracted jet spectra. The values listed in Table~\ref{tbl.fits}
are obtained from the un-subtracted jet spectra. The calibration of
\citet{warren14} was used to determine the densities, and we note that
the \citet{dz-cal} calibration would lead to marginally higher
values, while the original laboratory calibration leads to values
about 0.2--0.3~dex lower.  Comparing to the
background density value, there is no indication that the jet has an
enhanced density over the background. Atomic data for computing the
densities were obtained from version~7.1 of the CHIANTI atomic
database \citep{dere97,chianti71}. 

The only lines that retained a measurable intensity after background
subtraction had been performed were \lam195.12 and the \ion{Fe}{xi}
lines at 188.22 and 188.30~\AA\ that are blended. We can use the ratio
of \ion{Fe}{xii} to \ion{Fe}{xi} to derive a temperature, and the
values are given in Table~\ref{tbl.fits}. Atomic data were taken from
the CHIANTI atomic
database. It can be seen that the jet
plasma is a little cooler than the background plasma, but the
difference is small. It is clear that the jet plasma is not
appreciably hotter or cooler than the background coronal hole plasma.

\begin{deluxetable}{cccccc}
\tablecaption{Parameters for a dark jet and coronal hole background.\label{tbl.fits}}
\tablehead{
  Region &
  Intensity &
  Velocity &
  Width\tablenotemark{a} &
  $\log\,(N_{\rm e}/{\rm cm}^{-3})$ &
  $\log\,(T/{\rm K})$ \\
  & (\ecss) 
  & (\kms)
  & (m\AA)
  & 
}
\startdata
Background  & $14.7\pm 0.1$ 
            & 0 
            & $45\pm 3$ 
            & $9.33 \pm 0.07$ 
            & $6.13\pm 0.01$ \\
\noalign{\smallskip}
Region 1 & $6.5\pm 0.4$ 
         & $-52\pm 4$
         & $94\pm 6$ 
         & $8.85 \pm 0.10$ 
         & $6.11\pm 0.01$ \\
Region 2 & $2.8\pm 0.2$
         & $-83\pm 5$ 
         & $39\pm 7$ 
         & $9.06\pm 0.12$ 
         & $6.08\pm 0.02$ \\
Region 3 & $2.2\pm 0.2$ 
         & $-107\pm 5$
         & $54\pm 5$ 
         & $9.18\pm 0.08$
         & $6.11\pm 0.03$  \\
\enddata
\tablenotetext{a}{The instrumental line width has been subtracted.}
\end{deluxetable}

The mass flux in the jet can be compared with the typical proton mass
flux at 1~AU of 2--$4\times 10^8$~cm$^{-2}$~s$^{-1}$
\citep{wfeldman78}. Assuming a filling factor of 1 within the jet, the
particle flux 
is given by $0.85N_{\rm e}v$, where $N_{\rm e}$ is the electron number
density, 0.85 is the fraction of protons relative to electrons, and
$v$ the velocity along the jet axis. The heliocentric location of the jet bright point is
$(-455,-500)$, so if we assume the jet is perpendicular to the solar
surface, then the jet is inclined $45^\circ$ to the line-of-sight. If
we take a mean LOS velocity of 80~\kms\ (Table~\ref{tbl.fits}), then
this implies $v\approx 110$~\kms. The density of the jet plasma can
not be measured separately from the background plasma, but assuming
$\log\,N_{\rm e}=9$ gives a proton flux of $9.4\times
10^{15}$~cm$^{-2}$~s$^{-1}$. The EIS data do not allow accurate estimates of
jet lifetimes or frequencies, but the number of dark jets (11) is
comparable to the number of regular jets (13) in the HOP~177
data-set. If we assume the regular jets are the same as the X-ray jets
discussed by \citet{cirtain07}, then we expect 10 jets per hour, with
average lifetimes of 10~minutes \citep{savcheva07}. Assuming these
numbers apply to the dark jets, then we expect 1.7~jets on the Sun at
any instant. The cross-sectional area of the dark jet can be estimated
at 80~Mm$^{2}$ from the size of the structure in the EIS image
(Figure~\ref{fig.eg2}c). This then gives a proton flux at 1~AU estimate
of $2\times 10^{11}$~cm$^{-2}$~s$^{-1}$.
However, the very low intensity of the dark jet is not consistent with
the projected size and density. From an average intensity of
3.8~\ecss, a density of $\log\,N_{\rm e}=9$ and temperature
$\log\,T=6.1$ (Table~\ref{tbl.fits}) we can use the CHIANTI database to estimate a column
depth of the emitting plasma of only 100~km, compared to the projected
jet diameter of 5~Mm. This implies either a low filling factor of only
2\%\ or that the jet is actually a thin ``curtain'' of emission. The
proton flux is then modified to $4\times 10^9$~cm$^{-2}$~s$^{-1}$.
This value compares to $4\times 10^{11}$~cm$^{-2}$~s$^{-1}$ for
X-ray jets \citep{cirtain07}\footnote{The authors actually gave a
  value of $1\times 10^{16}$~cm$^{-2}$~s$^{-1}$, but we believe this
  is incorrect from the parameters tabulated in the paper.}, and
$1.2\times 10^{15}$~cm$^{-2}$~s$^{-1}$ from spicules \citep{athay82}. 

The dark jet mass flux estimate contains many assumptions, but it
suggests the mass flux is two orders of magnitude smaller than that
for regular coronal hole jets. To improve the dark jet estimate would
require (i) measurements of dark jet lifetimes
using high cadence spectral scans, and (ii) a determination of the
relative frequency of regular jets and dark jets.

\section{Summary}\label{sect.summary}

A continuous 2-day observation by the Hinode/EIS instrument of a large
coronal hole extension during 
2011 February 8--10 has revealed a number of jet events that are
identified only through a Doppler signature in the \ion{Fe}{xii}
\lam195.12 line (formed at 1.5~MK), with no counterpart in image
sequences obtained by the AIA instrument on board SDO. These jets are
named \emph{dark jets}. Of 24 jets
identified from the EIS data, 11 are classed as dark jets suggesting a
significant fraction of jet events are missed in surveys performed
with imaging instruments. The low intensity of the dark jets, however,
means that the total mass flux may be up to two orders of magnitude
lower than that from regular jets. 

The dark jet dopplergram images show a wide range of morphologies, but
the dark jets
are always associated with bright points, a feature in common with
regular jets. The lifetime of the dark jets cannot be constrained from EIS
data due to the low cadence of the rasters, but they do not live
longer than the 62~minute cadence of the EIS rasters.

An analysis of one dark jet observed by EIS on February 8 at 15:58~UT
revealed the following properties:
\begin{itemize}
\item The intensity enhancement of the dark jet plasma compared to the
  background plasma in the \ion{Fe}{xii} \lam195.12 line decreases
  from 44\%\ to 15\%\ along the length of the jet.
\item No evidence is found for a density enhancement compared to the
  background plasma, for which the density is $2.14\times
  10^9$~cm$^{-3}$.
\item The temperature of the dark jet plasma is 1.2--1.3~MK.
\item The LOS velocity of the dark jet plasma increases with height
  from $-52$ to $-107$~\kms, suggesting that acceleration continues
  along the jet axis.
\item The low jet intensity implies either a low filling factor (2\%)
  for the jet, or a curtain-like structure.
\end{itemize}
The properties of this dark jet are similar to those of the two
regular coronal jets presented by \citet{young14-pasj,young14-sp}. In
particular, these jets had temperatures of 1.3 and 1.7~MK,
respectively, and densities of 1.3 and 2.8 $\times$
$10^{8}$~cm$^{-3}$. In addition all three jets showed an increasing
speed with height, although the speeds were about a factor two lower
for the dark jet.
The implied curtain-like structure of the dark jet also matches that
of the \citet{young14-sp} jet, which had a much smaller line-of-sight width compared to the
plane-of-sky width. One difference is that regular coronal jets
generally correspond to a major change to the source bright point,
such as a strong brightening or morphology change. The bright points
underneath dark jets generally do not exhibit obvious changes
before or during the jet.
These facts suggest that dark jets may be triggered by a
smaller-scale, less-energetic mechanism than 
regular coronal jets, although the mechanism itself (such as magnetic
reconnection) may be the same, thus giving rise to similar plasma
parameters.

\acknowledgments
The author acknowledges funding from National Science Foundation grant
AGS-1159353, and thanks K.~Muglach for useful discussions.
The author thanks ISSI for financial support to attend the 2014
International Team Meeting ``Understanding Solar Jets and their Role
in Atmospheric Structure and Dynamics'' (PI: N.-E.~Raouafi), and he
thanks the participants for useful discussions.
\hinode\ is a Japanese mission developed and launched by 
ISAS/JAXA, with NAOJ as domestic partner and NASA and
STFC (UK) as international partners. It is operated by
these agencies in co-operation with ESA and NSC (Norway). SDO is a
mission for NASA's Living With a Star 
program, and data are provided courtesy of NASA/SDO and the AIA and
HMI science teams.

{\it Facilities:} \facility{Hinode(EIS)}, \facility{SDO(AIA)}

\bibliographystyle{apj}
\bibliography{myrefs}

\end{document}